          [2001/04/25 1.1 (PWD)]
\documentclass[a4paper,twocolumn]{esapub} 
\usepackage{natbib}
\usepackage{epsfig}

\newcommand{\tinyfigwidth}{8.5cm}

\title{Systematic Search for Short-transients and Pulsation Events 
from INTEGRAL Survey Data}
\author{Ken Ebisawa, Peter Kretschmar, Nami Mowlavi, Ada Paizis, 
Nicolas Produit, Simon Shaw}
\affil{INTEGRAL Science Data Centre, Chemin d'\'Ecogia 16, 1290
  Versoix, Switzerland}
\author{Sandro Mereghetti, Diego G\"otz}
\affil{CNR-IASF,  via Bassini 15, I-20133 Milano, Italy}
\author{Stefan Larsson}
\affil{Stockholm Observatory, AlbaNova, SE-106 91 Stockholm, Sweden}
\author{Niels Joergen Westergaard}
\affil{Danish Space Research Institute,
Juliane Maries Vej 30, DK-2100 Copenhagen \O,Denmark}
\author{Sami Maisala}
\affil{Observatory,
P.O. Box 14, FIN-00014 University of Helsinki, Finland}
\author{R\"udiger Staubert}
\affil{Institut f\"ur Astronomie und Astrophysik, Astronomie, 
Sand 1, D-72076 T\"ubingen, Germany}

\begin{document}


\maketitle

\begin{abstract}
The imaging instruments on board INTEGRAL have wide fields of view
 and high time resolution.  Therefore, they are ideal
 instruments to search for pulsating sources and/or transient
 events.  We are systematically searching for pulsations
 and transient events from known and
 serendipitous sources in the Galactic Plane Scan (GPS) and 
Galactic Center Deep Exposure (GCDE) core program
 data.  We analyze the standard pipe-line data using 
 ISDC Off-line Science Analysis (OSA) system, so that our results
 are reproducible by general guest users.   In this paper, we describe our system and
 report preliminary results for the first year of operation.
\end{abstract}

\begin{figure*}[htbp]
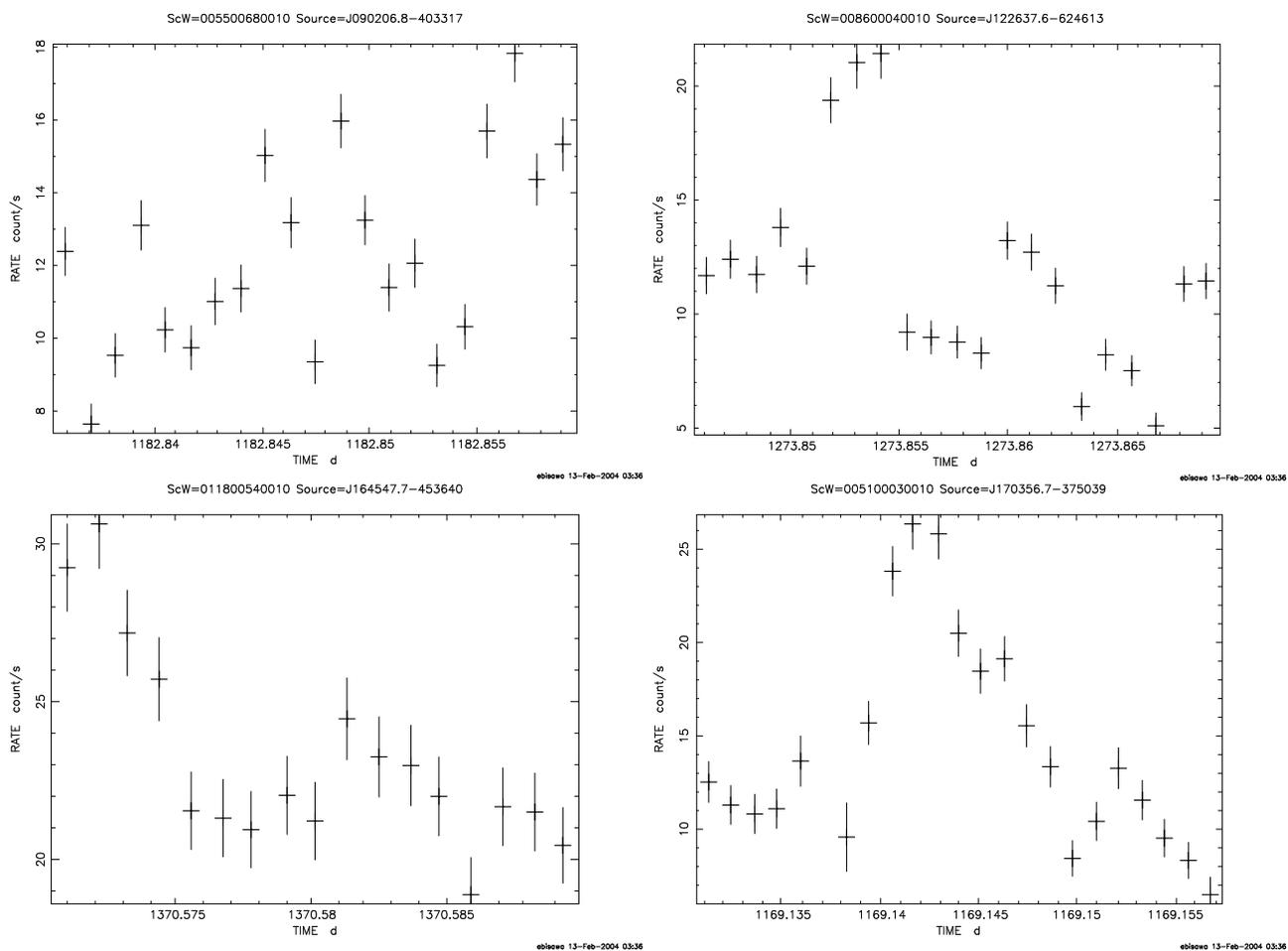

\centerline{
\epsfig{figure=J090206.8-403317_005500680010.eps,angle=270,width=\tinyfigwidth}
\epsfig{figure=J122637.6-624613_008600040010.ps,angle=270,width=\tinyfigwidth}
}
\centerline{
\epsfig{figure=J164547.7-453640_011800540010.ps,angle=270,width=\tinyfigwidth}
\epsfig{figure=J170356.7-375039_005100030010.ps,angle=270,width=\tinyfigwidth}
}
\caption{
Example of JEM-X light curves of variable sources, which show significant
variations within  a single Science Window.  Bin width is 100 seconds,
and the light curves are shown for each Science Window.
}
\end{figure*}

\begin{figure*}[htbp]
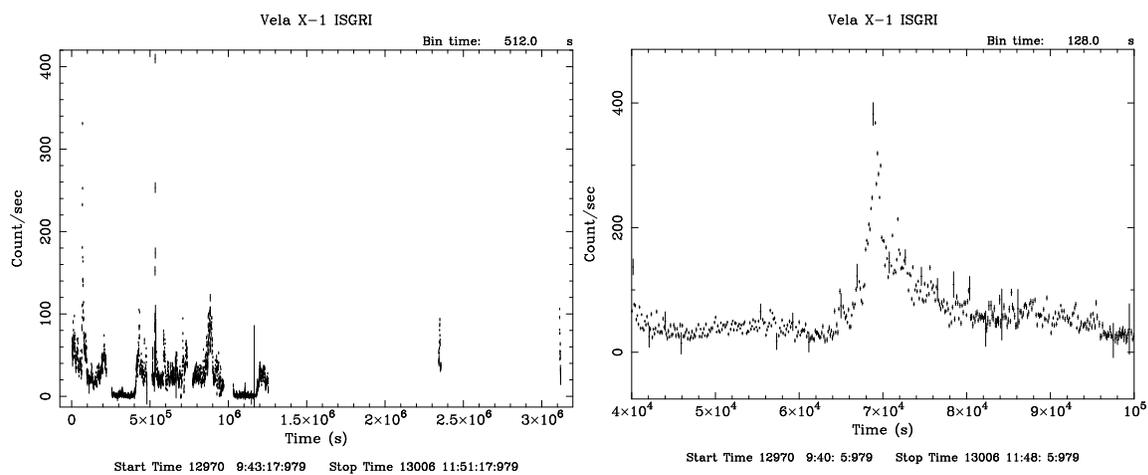

\centerline{
\epsfig{figure=vela_isgri_long.ps,angle=270,width=7.5cm}
\epsfig{figure=vela_isgri_short.ps,angle=270,width=7.5cm}
}
\caption{
Left: ISGRI Vela X-1 light curve from TJD = 12970 to 13006
(where Truncated Julian Date =  MJD - 40000).  
Bin-width is 512 sec, and the energy range is from 20 keV to 60 keV.
Right: Expansion of the large flare at around TJD =12971.  
Bin-width = 128 sec.
}
\end{figure*}

\section {Scientific Objective}
The wide fields of view and the dithering strategy of INTEGRAL observations enable us to carry out a systematic survey of a large area 
of the sky. The INTEGRAL Science Data Center (ISDC; Courvoisier et al.\ 2003), monitors the observations of the instruments with the {\it Quick Look Analysis} (QLA) software.  This allows us to systematically search for  bright transient events, as well as monitoring the primary targets of the observation,  within a few hours of receiving the telemetry (Shaw et al.\ 2004). 

However, the QLA carries out analysis only on the Science Window (ScW) basis.
Each ScW lasts about 30 minutes, and analysis at shorter timescales,  
which would take too much CPU time and disk space,  is not feasible with QLA. Therefore, we may miss short transient events, which are likely to be smeared if integrated for the entire ScW.
Also, QLA  does not search for pulsations, 
and fast pulsars down to $\sim$msec periodicity  may not be
detectable unless  event arrival time analysis is performed; 
such an analysis is not considered in the standard analysis pipe-line.

The INTEGRAL Burst Alert System (IBAS; Mereghetti 2003) produces
 real-time results of bright transient events; not only
Gamma-ray bursts, but also other transient events at various timescales
from milliseconds to minutes.  However, 
IBAS directly accesses raw telemetry, and  is not able to 
read the standard processed data. Also IBAS is not available to public.

To supplement the QLA,  standard analysis and IBAS, we are
developing a system to  systematically search for transient events and pulsations from 
core program GPS and GCDE data. Our system will be publicly available 
as a part of ISDC OSA.
In this paper, we describe the 
technical aspects of our project, and demonstrate
the feasibility.

\section {Analysis Methods}

\subsection{Transient events from known sources}

We routinely run the standard pipeline analysis on the GPS and GCDE data.
For the known and detected sources, we make ISGRI and JEM-X light curves
with 100 second time-bin\footnote{We use ISDC executables ``ii\_light'' 
and ``j\_src\_lc'' to create 
ISGRI and JEMX source light curves, respectively.}.

After the JEM-X and ISGRI light curves are made, we carry out 
statistical tests to search for variable events.  In Figure 1, we 
show examples of JEM-X light curves
with significant variations within one ScW.  
In this case, the 
search for variation is made for each ScW.  Namely, we 
find only variations shorter than the length of a ScW.  

Since each Scw is a single, stable pointing, these short-term variations are free from systematic effects that may be introduced by the observation dithering pattern.

In Figure 2, we show an example of the ISGRI Vela X-1 light curve. 
A prominent flare is seen at around TJD (Truncated Julian Date) = 12971 (Krivonos 
et al. 1994; Staubert et al. 2004).

\subsection{New transient search}

While IBAS systematically searches for strong gamma-ray bursts with IBIS,
soft X-ray transients below $\sim$15 keV may be expected only by JEMX. Also,
IBAS may not trigger weak gamma-ray bursts, which may be found in the
off-line analysis.

To search for new short-term transients, which are not in the
reference catalog, we need to carry out image deconvolution
and do a source search for a much shorter duration than the ScW.
Current OSA
 (v.3.1) is not really suitable for such a purpose,
since  a  user Good Time Interval (GTI) has to be specified for each Observation Group {\em before}\/ data correction and instrumental GTI calculation are made.

We adopt a new, more efficient data processing flow, which will be
standard in the INTEGRAL archives and OSA 4.0 (Figure 3).
In the new scheme, after the standard RAW, PRP, COR and GTI process,
a merged event list is created for each ScW.  Each event has
physical arrival time (TIME column), as well as START and STOP times of  GTI.
The GTI extensions are included in the same file, therefore generic tools
(such as xronos ftools, xselect) can read these event files.

After the event list is made for a ScW (which is typically
30 minutes or so), we define many short user GTIs, each 1 minute or so.
By applying the short user GTI, we run OSA and 
create sky images for  each $\sim$ 1 minute, and carry out a source search.
Note that in this method the user GTI is specified {\it after} the Correction
and instrumental GTI calculation step, so that we do not have to duplicate
these processes for each user GTI.

We have already adapted OSA so that it runs on the merged event file, and have 
started to systematically search for new transient events in the GPS and GCDE data.
We are evaluating sensitivity of ISGRI and JEMX for short transient events.
Preliminary results suggest we may detect $\sim50$ mCrab transient source for
one minute interval with both JEMX (5 -- 30 keV) and ISGRI (20 -- 60 keV).

\begin{figure*}[htbp]
\centerline{
\epsfig{figure=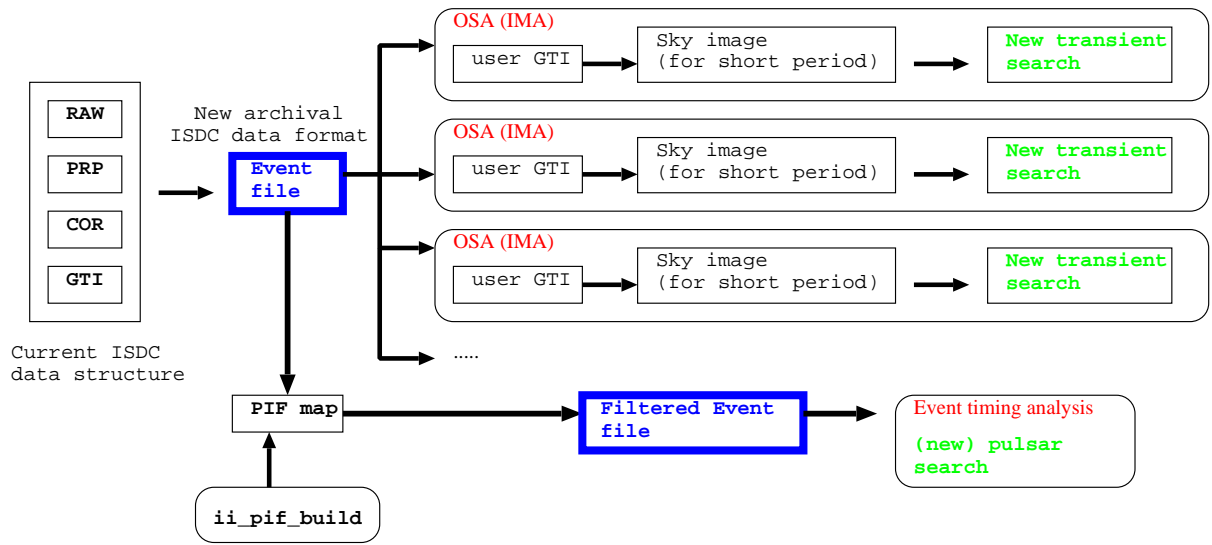,angle=270,width=16cm}
}
\caption{
Analysis flow for our short transient search and pulsar search program.
}
\end{figure*}

\begin{figure*}[htbp]
\centerline{
\includegraphics[bb=150 50 400 300,width=5.5cm,clip=true]{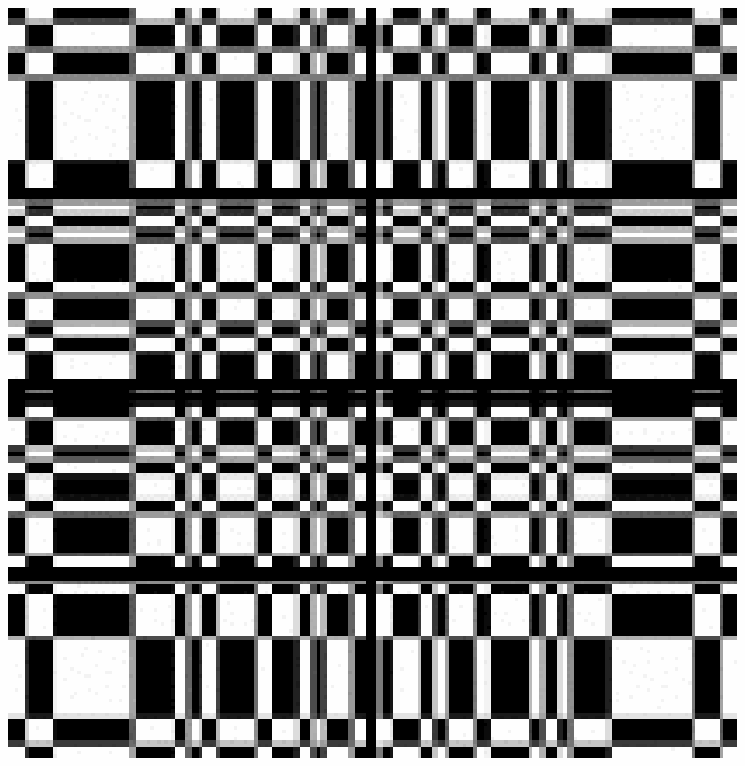} 
\includegraphics[bb=150 50 400 300,width=5.5cm,clip=true]{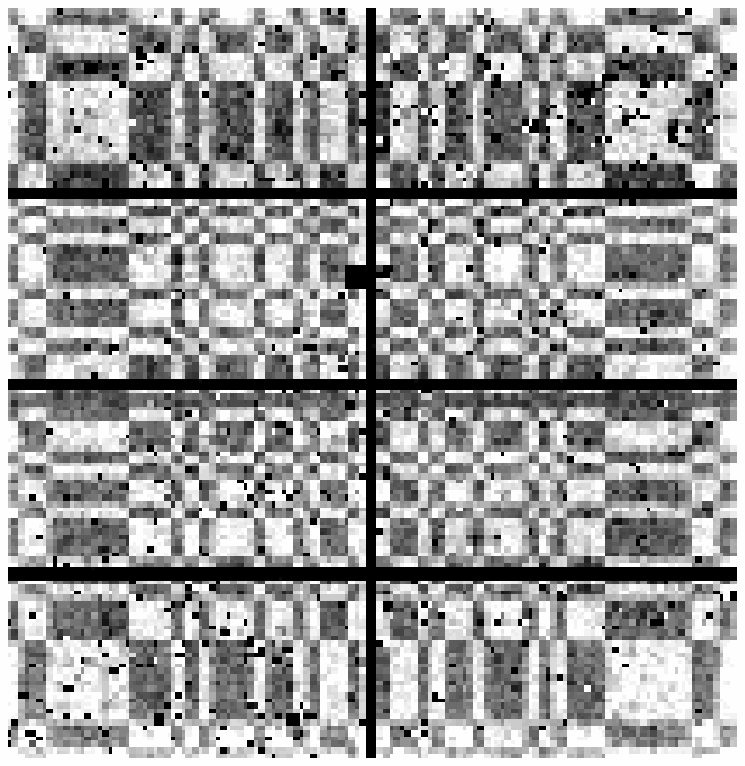} 
\includegraphics[bb=150 50 400 300,width=5.5cm,clip=true]{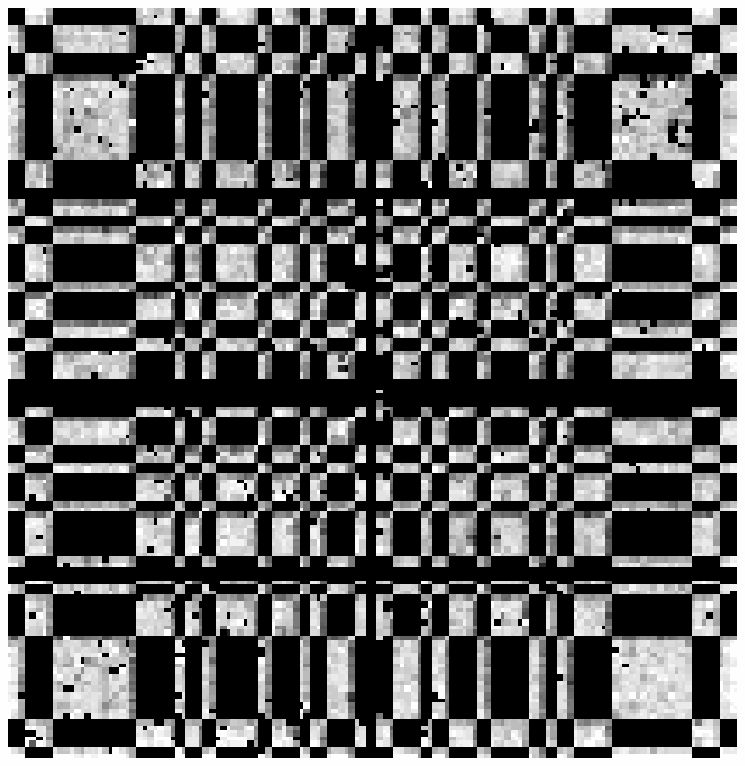} 
}
\caption{Left: ISGRI pixel Illumination Factor (PIF).  
Middle: Original detector image.  Right: Detector image
with pixels whose PIF values exceed 0.5. The data is from Science 
Window 010200380060 (Crab observation).}
\end{figure*}

\begin{figure*}[htbp]
\centerline{
\epsfig{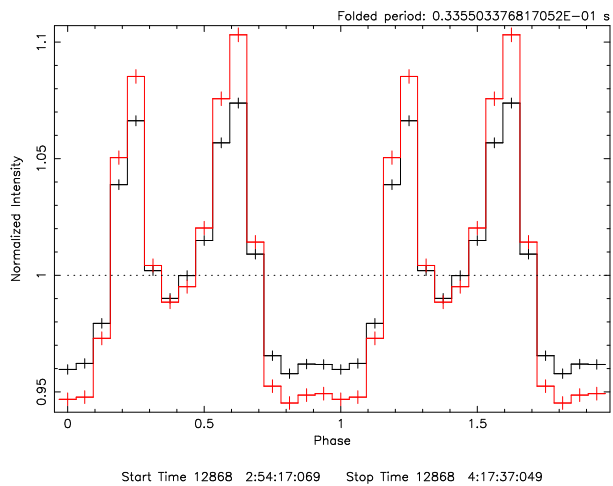}
}
\caption{Folded pulse-profile of the Crab pulsar from ISGRI 
ScW 010200380060. Black is made from the original event file 
using all the events, and red is from the events whose
PIF values exceed 0.5. 
Each profile is normalized so that the pulse-fractions
relative to the average counting rates are plotted.
}
\end{figure*}

\subsection{Short pulsation search}
To search for fast (e.g. $\sim $ millisecond) pulsations, we need to make event arrival time analysis without image deconvolution, since
it is not practical to carry out deconvolution for such a short period.
The ISDC tool ``barycent'' (since version 2.1) has been  modified so that it can 
perform the barycentric correction on the event lists.
To optimize the search for pulsations, we select events based 
on the Pixel Illumination Factor (PIF), which is the fractional area of each ISGRI pixel exposed to a source.  The PIF takes values from 0 to 1, such that PIF=0 means this pixel is not sensitive
to that source, and PIF=1 means this pixel is fully exposed to the
source.

A standard  and stand-alone tool {\tt ii\_pif\_build} is used to calculate the ISGRI PIF for a given source.  Once the PIF map is made, it is straightforward
to  filter event lists based on the PIF threshold (Figure 3).
In Figure 4, we show an example of the PIF and PIF-based event filtering.
By applying the PIF filter, we can reduce the background and optimize the
signal to noise ratio.  From experience, we choose PIF $>0.5$ as the event
selection criterion.  A similar PIF creation tool is expected for JEMX too.

In Figure 5, we demonstrate the effect of PIF selection.  In this example,
we analyze ISGRI Crab pulsar data from ScW 010200380060.
The events are folded with the apparent pulse-period (barycentric correction
is not made) at 33.550 msec.  We can see that, after the PIF selection, 
the background is substantially reduced and
the pulse-shape is much more prominent. Note that the {\tt fmaskfilt}
ftool is used to carry out the ISGRI PIF selection.

We have confirmed our event timing analysis method by detecting pulsations from 
 known pulsars such as Cen X-3.  By combining JEMX and ISGRI, we will
be able to study  known pulsars and search for new pulsars
in the wide energy range from $\sim5$ keV to $\sim200$ keV.

\section*{Acknowledgments}

We acknowledge the ISDC staff for the smooth operation of the satellite
and software supports.


\end{document}